\documentclass[pdflatex,sn-mathphys-num]{sn-jnl}

\usepackage{graphicx}%
\usepackage{multirow}%
\usepackage{amsmath,amssymb,amsfonts}%
\usepackage{amsthm}%
\usepackage{parskip}
\usepackage{physics}
\usepackage{mathtools}
\usepackage{mathrsfs}%
\usepackage[title]{appendix}%
\usepackage{xcolor}%
\usepackage{textcomp}%
\usepackage{manyfoot}%
\usepackage{booktabs}%
\usepackage{algorithm}%
\usepackage{algorithmicx}%
\usepackage{algpseudocode}%
\usepackage{listings}%
\usepackage{lscape}
\usepackage{siunitx}

\theoremstyle{thmstyleone}%
\theoremstyle{thmstyletwo}%

\theoremstyle{thmstylethree}%

\begin{document}
\title[Article Title]{$M^2$ as a Quantitative Measure for Beam Quality}


\author*[1,2]{\fnm{Filipp} \sur{Lausch}}\email{filipp.lausch@univie.ac.at}

\author[1]{\fnm{Vito} \sur{F. Pecile}}\email{vito.pecile@univie.ac.at}

\author[1]{\fnm{Oliver} \sur{H. Heckl}}\email{oliver.heckl@univie.ac.at}

\affil[1]{\orgdiv{University of Vienna}, \orgname{Optical Metrology Group}, \orgaddress{\street{Boltzmangasse 5}, \postcode{1090}, \state{Vienna}, \country{Austria}}}

\affil[2]{\orgdiv{University of Vienna}, \orgname{Vienna Doctoral School in Physics}, \orgaddress{\street{Boltzmangasse 5}, \postcode{1090}, \state{Vienna}, \country{Austria}}}


\abstract{
Beam quality is a fundamental aspect for evaluating the performance of laser sources. $M^2$-measurements serve as the gold standard for beam quality assessment since the 1990s. The measured $M^2$-parameter indicates similarity to the pure fundamental Gaussian mode, characterized by the ideal $M^2=1$, by describing a beams' divergence. $M^2$-values close to 1 are considered to correspond to nearly fundamental sources. However, in terms of the higher-order mode contribution of a laser, it acts as a qualitative measure that does not permit a quantitative statement. Here, we introduce a framework to assess the fundamental mode content of a laser beam using $M^2$-measurements and establish a direct link between beam quality and its mode composition. Our results significantly enhance the utility of $M^2$-measurements in evaluating laser sources, coupling efficiencies, focusing performance, and long-distance propagation. This repositions $M^2$ from a qualitative figure to a quantitative tool in modern photonics.}




\maketitle

\section{Introduction}
For over thirty years, the $M^2$ parameter has been the standard for spatial beam characterization. Introduced by A.E. Siegman in 1990 \cite{Siegman:1990} and further developed in subsequent works \cite{Siegman:1992, Siegman:19922, Siegman:1993, Siegman:1998}, it is based on the $\mathrm{D}4\sigma$-definition of beam width and provides a framework for describing spatial beam quality for arbitrary superpositions of Hermite-Gauss modes, including the fundamental mode and higher-order modes (HOMs). Today, this approach is widely used in both academia and industry and has been standardized in ISO 11146-1 \cite{ISO:2021}.

In most applications, such as recently for our optical parametric oscillator as part of a mid-infrared light source for cavity-enhanced Lamb-dip spectroscopy \cite{Heckl:2024,Pecile:2024}, a clean fundamental Gaussian mode ($\text{TEM}_{00}$) is critical for achieving tight focusing, long-distance beam collimation, or efficient coupling to resonators and optical fibers. Smaller $M^2$ values indicate higher beam quality, with $M^2 = 1$ corresponding to an ideal fundamental Gaussian mode. However, while it is known that smaller $M^2$ values correlate with a larger fraction of the fundamental mode, no mathematical expression exists which can quantify the contributions of individual modes, making an accurate estimation of the fundamental mode contribution impossible.

In this work, we overcome this limitation by establishing an extensive relationship that transforms an $M^2$-measurement into a direct power estimator. On one hand we derive a generally applicable lower bound to the fundamental modes' relative power $P_{00}$, and on the other hand we introduce a low-$M^2$ ($M_x^2+M_y^2\leq 4$) approximation for the upper bound, such that
\begin{align}
   2-\frac{M_x^2+M_y^2}{2} \leq P_{00}\leq 1.5-\frac{M_x^2+M_y^2}{4}\,.\label{awesome0}
\end{align}
This fundamentally advances the applicability of $M^2$-measurements by linking beam divergence to fundamental mode purity. The resulting metric quantifies a beam's $\text{TEM}_{00}$ content -- a critical parameter for efficient coupling to, e.g., optical resonators and fibers. It enables more precise characterization of light sources and holds significant relevance for atomic, molecular and optical physics, quantum optics, and various applications in the photonics industry.

\section{Preliminaries}
In this section we give a short review of the mathematical foundation of our considerations.
Throughout the manuscript, we regard the general case of a superposition of an arbitrary finite amount of transverse electromagnetic modes ($\text{TEM}_{mn}$) of a monochromatic Hermite-Gauss beam traveling along the z-axis\,\cite{Siegman:Lasers}. The indices $m, n \in \mathbb{N}_{0}$ are the orders of the mode in x- and y-direction, respectively. The electrical field amplitude $E(x,y,z)$ of the beam is constructed with coefficients $c_{mn}\in \mathbb{C}$, which allow us to mathematically switch specific modes on and off at our convenience:
\begin{align}
    E(x,y,z) &= e^{-ikz} \sum_{m=0}^M \sum_{n=0}^N c_{mn} u_{m}(x,z) u_{n}(y,z)\,,\label{efield0}\\
    u_m (x,z) &= \left( \frac{2}{\pi}\right)^{1/4}\sqrt{\frac{e^{-i(2m+1)(\psi (z)-\psi_0)}}{2^m m! \omega (z)}} 
    H_m\left(\frac{\sqrt{2}x}{\omega (z)}\right) e^{-\frac{ikx^2}{2R(z)}-\frac{x^2}{\omega^2 (z)}}\,.\label{efield}
\end{align}
$\psi (z)$ is the Gouy phase, $R(z)$ the radius of curvature, $k$ the wavenumber, and $w(z) = \omega_0\sqrt{1+z^2/z_R^2}$ half of the fundamental beam ($\text{TEM}_{00}$) width with Rayleigh length $z_R$ and minimal waist radius $w_0$. The functions $u_m(x,z)$ and $u_n(y,z)$, defined in the same way with $x\leftrightarrow y$, form a complete basis and are orthonormal \cite{Siegman:Lasers}:
\begin{align}
    \int_{-\infty}^\infty dx \, u^*_m(x)u_n(x)\, &= \delta_{mn}\,.\label{orth1}
\end{align}
The total cross-sectional power $P$ at an arbitrary position $z$ of this beam can be calculated via integration over the $x-y-$plane:
\begin{align}
    P &= \int_{-\infty}^\infty dx \int_{-\infty}^\infty dy \,I(x,y,z) =  \sum_{m=0}^M \sum_{n=0}^N
    \abs{c_{mn}}^2\,. \label{power}
\end{align}
The optical intensity $I(x,y,z)$ is given as
\begin{align}
    I(x,y,z) = \abs{E(x,y,z)}^2.
\end{align}
We can now rewrite and normalize equation \eqref{power}, which yields
\begin{align}
    P_{mn}\coloneqq\frac{\abs{c_{mn}}^2}{P}\in [0,1]\,, \label{rel_pwr}
\end{align}
where $P_{mn}$ is the relative power contribution of an arbitrary mode $\text{TEM}_{mn}$ to the total cross-sectional power of the investigated beam.

In the following, we now establish the mathematical basis that underlies $M^2$ measurements. Note, that we closely follow the notation used by Siegman in his original works\,\cite{Siegman:1990,Siegman:1992,Siegman:19922,Siegman:1993,Siegman:1998}. The width of an arbitrary beam is based on the $\mathrm{D4\sigma}$ or second moment definition and denoted as $W_x(z)$ and $W_y(z)$ in x- and y-directions, respectively. It is defined as
\begin{align}
    W_x^2(z) &=\frac{4\int dxdy \,I(x,y,z)(x-\overline{x})^2}{\int dxdy \,I(x,y,z)}\,,\label{1}\\
    W_y^2(z) &=\frac{4\int dxdy \,I(x,y,z)(y-\overline{y})^2}{\int dxdy \,I(x,y,z)}\,,\label{2}
\end{align}
with $\overline{x}$ and $\overline{y}$ being the beam centroids in their respective directions.

Without loss of generality, we can assume that the beam centroid is given by $\overline{x}=\overline{y}=0$. Experimentally, this can usually be achieved within the uncertainty of the measurement device, which we assume to be negligible.

To receive the $M^2$ value of a real laser beam experimentally, a set of beam widths at different $z$ positions is obtained using the width definitions from equations \eqref{1} and \eqref{2}.
The beam parameters $M_x^2$ and $M_y^2$ (indexed corresponding to their axis of measurement) are then  retrieved by fitting the measured $W_x(z)$ and $W_y(z)$ data to 
\begin{align}
    W^2_{x}(z) &= W_{0x}^2 + \left(\frac{\lambda M_{x}^2}{\pi W_{0x}}\right)^2 (z - z_{0x})^2\,,\label{qudfit0}\\
    W^2_{y}(z) &= W_{0y}^2 + \left(\frac{\lambda M_{y}^2}{\pi W_{0y}}\right)^2 (z - z_{0y})^2\,,\label{qudfit}
\end{align}
with $\lambda$ being the wavelength and $W_0$ the minimal beam waist at $z=z_0$, indexed corresponding to the respective axis of measurement\,\cite{Siegman:1990}. Without loss of generality, we define $z_{0x}=z_{0y}=0$.

Calculating equations \eqref{1} and \eqref{2} for arbitrary transversal mode superpositions of monochromatic Hermite-Gauss beams, as defined in equation \eqref{efield0}, always yields an expression that is quadratic in $z$. As a result, it is possible to correlate the experimentally found $M^2$-parameters with the theoretical second order coefficients in $z$.

Now we have deployed the necessary mathematical background to formulate the problem of this work. We plug equations \eqref{efield0} and \eqref{efield} into equations \eqref{1} and \eqref{2}. Next, we link equations \eqref{1} and  \eqref{2} with equations \eqref{qudfit0} and \eqref{qudfit}, respectively. We find the following equations:
\begin{align}
 \omega^2 \cdot \sum_{m=0}^{M}\sum_{n=0}^{N} (2m+1)P_{mn} &= W_{0x}^2 + \left( \frac{M_x^2\lambda}{\pi W_{0x}}\right)^2 z^2\,,\label{eqr1}\\
\omega^2 \cdot \sum_{m=0}^{M}\sum_{n=0}^{N} (2n+1)P_{mn} &= W_{0y}^2 + \left( \frac{M_y^2\lambda}{\pi W_{0y}}\right)^2 z^2\,.\label{eqr2}
\end{align}
Using the condition of $M_y^2/M_x^2 = W_{0y}^2/W_{0x}^2$, equations \eqref{eqr1} and \eqref{eqr2} can be simplified\,\cite{Siegman:1990,Siegman:1998}. Together with the normalized cross-sectional power of the Hermite-Gauss beam (inserting equation \eqref{rel_pwr} into equation \eqref{power}) this yields the following system of three equations:
\begin{align}
\sum_{m=0}^{M}\sum_{n=0}^{N} (2m+1)P_{mn} &= M_x^2\label{sol1}\,,\\
\sum_{m=0}^{M}\sum_{n=0}^{N} (2n+1)P_{mn} &= M_y^2\,,\label{sol2}\\
\sum_{m=0}^{M}\sum_{n=0}^{N} P_{mn} = 1\,.\label{sol3}
\end{align}
These equations form the foundation of our subsequent analysis. Equations \eqref{sol1} and \eqref{sol2} were already described by A. E. Siegman in 1990\,\cite{Siegman:1990}, opening up the world of $M^2$-measurements not only for fundamental Gaussian beams, but also for beams composed of an arbitrary superposition of higher-order modes.

\section{Results}
\subsection{Lower Bounds}\label{lowbou}
For simplicity, we first assume that equations \eqref{sol1}-\eqref{sol3} form a system of equations that eliminates precisely 3 of the $M\cdot N$ degrees of freedom given through the $P_{mn}$. This will allow us to calculate the most general case for the fundamental mode power estimation. This estimate is based on a specific type of parametrization that does not cover all possible beams. The corresponding special cases will be provided below.

For any combination of three $P_{mn}$ a parametrization in dependence of the remaining $M\cdot N-3$ parameters can be derived. We are specifically interested in parametrizations involving $P_{00}$. While it is not possible to find a general Ansatz that represents every single possible case, we can make one that generalizes as much as possible. Given two modes $\text{TEM}_{ab}$ and $\text{TEM}_{cd}$ with $bc-ad\neq 0$, the following parametrization achieves this:
\begin{align}
    P_{ab} =& \frac{d(M_x^2-1) - c(M_y^2-1)}{2(ad-bc)}  +\sum_{\substack{(m,n)\neq (a,b);\\(c,d)}} \frac{cn-dm}{ad-bc}P_{mn},\label{Param1}\\
    P_{cd} =& \frac{a(M_y^2-1) - b(M_x^2-1)}{2(ad-bc)}   -\sum_{\substack{(m,n)\neq (a,b);\\(c,d)}} \frac{an-bm}{ad-bc} P_{mn}\, , \\
    P_{00} =& 1 -\frac{(c-a)(M_y^2-1)+(b-d)(M_x^2-1)}{2(bc-ad)} \nonumber\\&+ \sum_{\substack{(m,n)\neq (0,0);\\(a,b);(c,d)}}\left(\frac{m(b-d)+n(c-a)}{bc-ad}-1\right)P_{mn}\,.\label{Param3}
\end{align}
For the fundamental mode power estimation, we are interested in the lower bounds of this parametrization. Generally, due to the possibility of $(m(b-d)+n(c-a))/(bc-ad)< 1$, determining the minimum of $P_{00}$ requires extensive analysis of the case-specific parameter-space. However, for a general estimate it turns out to be sufficient to analyze the parametrizations with $(m(b-d)+n(c-a))/(bc-ad)\geq 1$ $\forall m,n$. In this case all coefficients in equation \eqref{Param3} are positive, hence yielding a lower bound whenever the free parameters are set to zero (i.e. $P_{mn}=0$ $\forall m,n$):
\begin{align}
     P_{00}\geq & 1 -\frac{(c-a)(M_y^2-1)+(b-d)(M_x^2-1)}{2(bc-ad)}\,.\label{ineq1}
\end{align}
The reason for the sufficiency of this simplification is given through analyzing the upper bounds of the higher-order modes. Because of the coefficients $2m+1$ and $2n+1$ in equations \eqref{sol1} and \eqref{sol2}, higher-order modes are increasingly more limited in their upper bound, i.e. $P_{mn}\leq \text{min}\lbrace (M_x^2-1)/(2m),(M_y^2-1)/(2n),1\rbrace$. $\text{TEM}_{10}$ and $\text{TEM}_{01}$ yield the maxima of those upper bounds through $P_{10}\leq \text{min}\lbrace (M_x^2-1)/2,1\rbrace$ and $P_{01}\leq \text{min}\lbrace (M_y^2-1)/2,1\rbrace$. As a result, the minimum of all possible lower bounds to $P_{00}$ is limited by the case where a parametrization involving both modes is chosen and the lower bound computed. As it turns out, inequality \eqref{ineq1} covers this scenario for $b=c=0$ and $a=d=1$ with positive coefficients $m+n-1\geq 0$ $\forall m,n$ for the free parameters in equation \eqref{Param3}.

Realistically, it is challenging to have precise information about the onset of HOMs. As a result, we have to assume a parametrization with the possible presence of all HOMs. This is in agreement with the result of a parametrization involving $\text{TEM}_{01}$ and $\text{TEM}_{10}$ yielding the minimal lower bound to $P_{00}$. We receive the general expression for estimating the lower bound of the fundamental mode's power in presence of HOMs:
\begin{align}
    \boxed{P_{00} \geq 2 - \frac{M_x^2+M_y^2}{2}}\,.\label{awesome}
\end{align}

In Fig. 1 we demonstrate our findings for an example superposition of four different TEMs. For a beam with known composition the full range of $P_{00}$ can be determined, as can be seen in the corresponding plots of the parametrization. In section \ref{sec32} we find an approximation to the full range of $P_{00}$ specifically aimed at low-$M^2$ systems with $M_x^2+M_y^2\leq 4$.
\begin{center}
\includegraphics[scale=0.3]{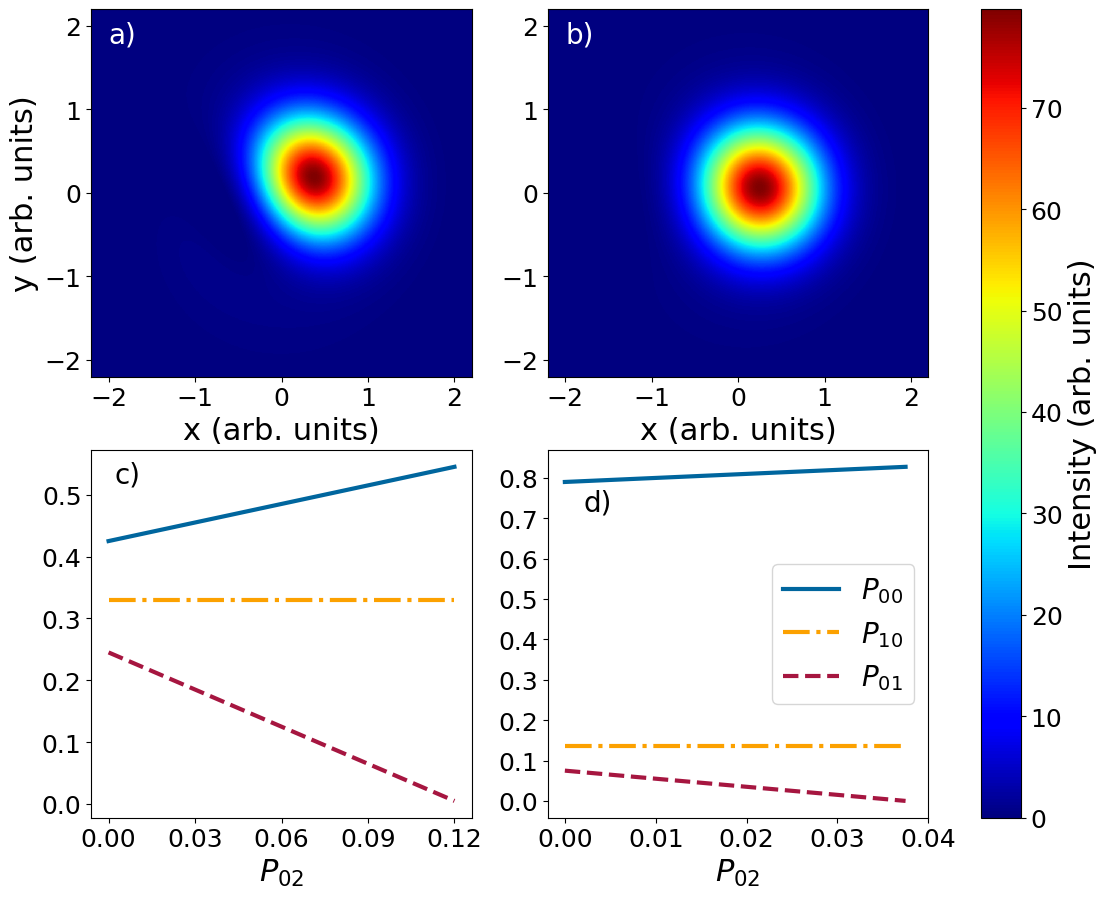}
\end{center}
\small{Fig. 1: Superpositions of $\text{TEM}_{00}$, $\text{TEM}_{10}$, $\text{TEM}_{01}$ and $\text{TEM}_{02}$ with a) $M_x^2 = 1.66$ and $M_y^2 = 1.49$ as well as b) $M_x^2 = 1.27$ and $M_y^2=1.15$. The corresponding lower bounds: a) $P_{00} \geq 0.425$ and b) $P_{00} \geq 0.79$. Plot c) (resp. d)) is the plot of the parametrization of a) (resp. b)) for the given $M^2$-values. For small numbers of HOMs it is easily possible to analyse the full range of $P_{00}$, as can be seen in c) and d). The parametrizations not only yield a lower bound, but also an upper bound such that $0.425 \leq P_{00}\leq 0.5475$ for scenario a) and $0.79 \leq P_{00}\leq 0.8275$ for scenario b). $P_{02}$ being limited through $P_{01}\geq 0$ enables the calculation of upper bounds to $P_{00}$.}

An interesting question that arises from inequality \eqref{awesome} is whether or not it is possible to determine allowed $M^2$-values for a given relative power of the fundamental mode $P_{00}$. As long as the fundamental mode only has to fulfill that its relative power is at least $P_{00}$, the following inequality describes all corresponding allowed pairs $(M_x^2,M_y^2)$:
\begin{align}
    M_x^2+M_y^2\leq 4-2P_{00}\label{awesome2}\,.
\end{align}
We note that the applicability of this inequality is not as general as inequality \eqref{awesome}, as the possible $M^2$-values are strongly limited. Hence, its application will play a role whenever both high fundamental mode power as well as low $M^2$-parameters are at hand. For more general statements, the case specific lower bounds have to be derived.
\begin{center}
\includegraphics[scale=0.4]{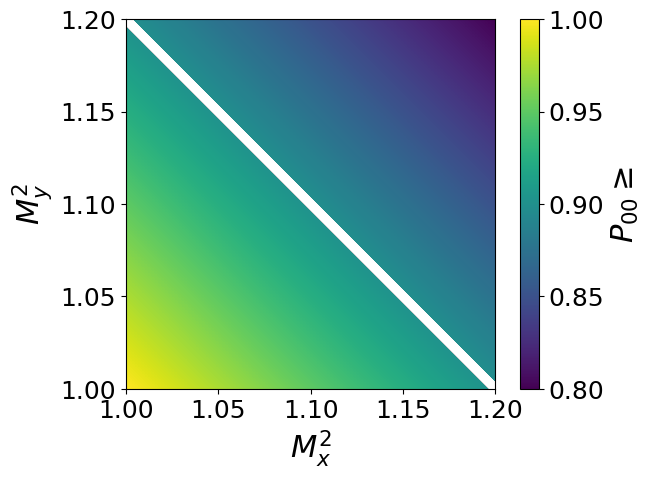}
\end{center}
\small{Fig. 2: Graphic representation of the solutions to inequality \eqref{awesome2}. Every point in the heat map on and below the white line corresponds to a combination of $M^2$-values of a beam for which the relative power of the fundamental mode is higher than or equal to 0.9.}

\subsubsection{Special Cases}
Some solutions of the equations \eqref{sol1}-\eqref{sol3} are not contained within inequality \eqref{ineq1}.  Here, we complete the picture by discussing their symmetry characteristics as well as providing the lower bounds for the fundamental mode. Whenever the superposition of higher-order modes shows some type of symmetry, the system of equations simplifies.

We first comment on the special case of $bc-ad=0$. Here, either $P_{ab}$ or $P_{cd}$ cancels out through a coefficient containing $bc-ad$ during the process of calculating the parametrization. Hence, precisely one different $P_{mn}$ has to be parametrized instead of one of the other two. Choosing $P_{ef}$, such that $af-be\neq 0$, one reproduces equation \eqref{Param3} with the replacements $c\rightarrow e$ and $d\rightarrow f$
which leads to the same conclusion as in inequality \eqref{awesome} under the assumption of positive coefficients for the free parameters. The difference is the independence of $P_{ef}$ of the free parameter $P_{cd}$.

Following up, for the remaining special cases all lower bounds are of larger or equal magnitude compared to inequality \eqref{awesome}. Consequently, inequality \eqref{awesome} can be used for any beam with the downside that it might underestimate the fundamental mode. In general, the upcoming symmetries are so specific, that this underestimation will rarely be the case. We list them with respect to their symmetry characteristics:
\begin{enumerate}
\begin{small}
\item \textbf{Sums of $\text{TEM}_{00}$ and $\text{TEM}_{m_in_i}$ with $m_i/n_i=\alpha\neq 0$ $\forall i$, $a=\text{min}(\lbrace m_i\rbrace)$ and $\alpha \in \mathbb{Q}$:}
\begin{align}
    P_{00} &\geq \frac{2a+1-M_x^2}{2a}\,, \quad M_x^2 = \alpha (M_y^2-1) + 1\,.\label{spe1}
\end{align}
This class of beams entails e.g., sums of $\text{TEM}_{nn}$ with $\text{TEM}_{aa}$ as lowest HOM. Generally, sums of $\text{TEM}_{m_in_i}$ with $m_i=ia$ and $n_i=ib$, such that $ia, ib\in \mathbb{N}_0$ $\forall i$ as well as $i \in \mathbb{N}_0$ and $a/b\neq 0$ yield this kind of lower bound. The case $\alpha=0$ corresponds to sums of $\text{TEM}_{m0}$ and sums of $\text{TEM}_{0n}$. Moreover, cases 1.-3. correspond to all differentiations for sums that fulfill $m_in_j-m_jn_i=0$ $\forall i,j$.
\item \textbf{Sums of $\text{TEM}_{m0}$ with $\text{TEM}_{a0}$ as lowest HOM:}
\begin{align}
    P_{00} \geq \frac{2a+1-M_x^2}{2a}, \qquad M_y^2 = 1.
\end{align}
 \item \textbf{Sums of $\text{TEM}_{0n}$ with $\text{TEM}_{0a}$ as lowest HOM:}
\begin{align}
    P_{00} \geq \frac{2a+1-M_y^2}{2a}, \qquad M_x^2 = 1.\label{spe3}
\end{align}   
\item \textbf{Sums of $\text{TEM}_{00}$ and $\text{TEM}_{m_in_i}$ such that $m_i < m_{i+1}$ and $n_i < n_{i+1}$ $\forall i$ with at least one pair of indices such that $m_in_{j}-m_{j}n_{i}\neq 0$ for $i\neq j$:}

This case admits parametrizations corresponding to the equations \eqref{Param1}-\eqref{Param3} with at least one negative coefficient for the free parameters that can not be removed through reparametrization. As a result, case-specific analysis of the parameter space always has to be applied. As such, the recommended approach is to assume inequality \eqref{awesome} in this scenario.
\end{small}
\end{enumerate}
We note that the cases given through sums of $\text{TEM}_{an}$ ($a=$const.; $a\geq 1$) and sums of $\text{TEM}_{ma}$ ($a=$const.; $a\geq 1$) are contained within the equations \eqref{Param1}-\eqref{Param3}, however, equation \eqref{Param3} is constrained to equality.

\subsection{Upper Bounds}\label{sec32}
As we already hinted in Fig. 1, for a small number of HOMs in a beams' composition, the full range of $P_{00}$ can easily be accessed. The derivation of upper bounds will in general be case-specific and thus no generally applicable expression can be derived. However, we can use the combination of small superpositions as well as low $M^2$ values to our advantage to find an approximation to the fundamental mode power upper bound that is applicable in a vast spectrum of applications, such as single-mode light sources, optical fibers as well as coupling into optical resonators.
 
First of all, it is of interest to analyze the behavior of the parametrizations for combinations of $M^2$-values near 1. We choose sufficiently small values fulfilling $M_x^2+M_y^2 \leq 4$ such that the generalized lower bound in inequality \eqref{awesome} fulfills $2-M_x^2/2-M_y^2/2\geq 0$. This simplifies the discussion as it prevents negative valued lower bounds, which in general might force an onset of HOMs and as a result always require a case-specific analysis of the parameter space. All of the parametric solutions are restricted by a boundary value problem, in its most general form given by
\begin{align}
    0\leq P_{ab}\leq 1\,,\label{bound1}\\
    0\leq P_{cd}\leq 1\,,\\
    0\leq P_{00}\leq 1\,.\label{bound3}
\end{align}
If we describe the solution of $P_{00}$ in the space spanned by $P_{00}$ and the set of free parameters $\lbrace P_{mn}\rbrace$, one finds that it corresponds to a hyperplane with well defined boundary. This hyperplane is given through the set of all points $\Sigma_{M_x^2,M_y^2} :=  \lbrace (P_{00}(\lbrace P_{mn}\rbrace),\lbrace P_{mn}\rbrace)\rbrace$. We index it with the corresponding $M^2$-parameters as they influence the size of the boundary $\partial\Sigma_{M_x^2,M_y^2}$ for a fixed set of modes within the mode expansion. A purely fundamental beam has $M_x^2=M_y^2=1$ and as a result $P_{00}=1$ whilst the relative power of all other modes has to be zero. This implies that the hyperplanes always have to converge to the point $\Sigma_{1,1} = \lbrace P_{00}(\lbrace P_{mn}\rbrace)=1\rbrace$ as $M_x^2$ and $M_y^2$ simultaneously tend to 1. Due to the linear dependence in the $M^2$-parameters, the convergence is also linear with respect to each coordinate-direction. This means that for high fundamental mode lower bounds near 1 we also find a stronger restriction of the upper bound of $P_{00}$ compared to smaller lower bounds. We demonstrate this in Fig. 3 for the simplified case of $M_x^2=M_y^2$ for the sake of representability.
\begin{center}
    \includegraphics[scale=0.5]{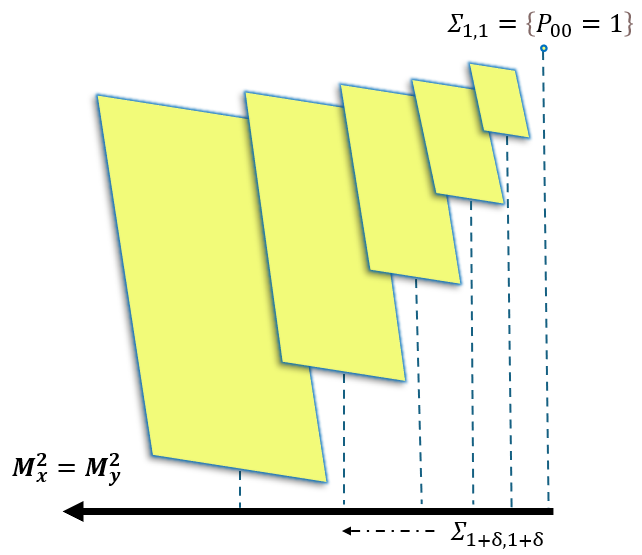}\\
\end{center}
\small{Fig. 3: Schematic representation of the convergence of the hyperplanes for the setting of $M_x^2=M_y^2$ and $\delta\leq 1$.}

The region $M_x^2+M_y^2\leq 4$ corresponds to $M^2$-values that fit high-precision applications like coupling to optical resonators, single mode fibers, mode-cleaning of high-power sources, etc. Single mode fibers exhibit low $V$-parameters of $V< 2.405$, only supporting at most one mode in each polarization direction \cite{ST:2007}. The $V$-parameter is a fiber parameter that determines the allowed number of modes inside of the fiber. Low-loss high-finesse optical resonators also support only a low number of $\text{TEM}_{mn}$-modes close to the fundamental mode, even yielding a mode-cleaning effect for high-power light sources. Furthermore, for low $M^2\approx 1$ HOMs will be strongly suppressed. Their relative power is at least bounded by $P_{mn}\leq \text{min}\lbrace(M_x^2-1)/(2m),(M_y^2-1)/(2n)\rbrace$. As a result, for the upper bound approximation we assume only HOMs of order $m+n\leq 2$ due to them having the most significant contribution. Importantly, we note that HOMs with the same order in the sum $m+n$ have one common intra-cavity resonance frequency \cite{ST:2007}, making their contributions hard to distinguish in practice.  $m+n\leq 2$ corresponds to a superposition of $\text{TEM}_{00}$ with $\text{TEM}_{10}$, $\text{TEM}_{01}$, $\text{TEM}_{11}$, $\text{TEM}_{20}$ and $\text{TEM}_{02}$ with the following parametrization:
\begin{align}
    P_{10}&=\frac{M_x^2-1}{2}-P_{11}-2P_{20}\,,\label{bref1}\\
     P_{01}&=\frac{M_y^2-1}{2}-P_{11}-2P_{02}\,,\label{bref2}\\
      P_{00}&=2-\frac{M_x^2+M_y^2}{2}+P_{11}+P_{20}+P_{02}\,.
\end{align}
From this, the upper bound of the fundamental mode power can be retrieved. It forms the following inequality together with the lower bound in inequality \eqref{awesome}:

\begin{align}
    \boxed{2-\frac{M_x^2+M_y^2}{2}\leq P_{00}\leq 1.5-\frac{M_x^2+M_y^2}{4}}\,.\label{awesome3}
\end{align}

It is now possible to accurately determine the purity of a light beam by applying the interval given by the inequality. Specifically at low $M^2$-values this accuracy increases linearly. This is highly relevant in a wide range of applications, from which we discuss the most relevant ones in the upcoming section \ref{sec4}.

\section{Discussion}
\subsection{Implications for real-world optical systems}\label{sec4}
So far we have mentioned coupling to optical resonators, single-mode light sources as well as single-mode fibers as relevant fields of application for the fundamental mode purity estimation. Here, we go more in-depth on the corresponding applicability and discuss examples.
\begin{itemize}  
\begin{small}
    \item \textbf{Optical fibers and fiber lasers (FLs):}\\
    Single-mode fibers and FLs based on rare-earth-doped gain-fibers usually have $M^2<1.1$ (see Table \ref{tab:my_label}). As a result, they yield $P_{00}\in [0.9,0.95]$ or better. Multi-mode fibers are associated with a wide range of possible $M^2$-values, ranging up to $M^2=4$ \cite{ST:2007}, but supporting HOMs, they can go far beyond that. Analyzing up to the non-negative boundary (negative lower bounds require case-studies) of the general lower bound in inequality \eqref{awesome} at $M_x^2+M_y^2=4$ with $P_{00}\geq 0$, one finds a range for the upper bounds given by the interval $[0,0.5]$. Extending the model with more higher-order mode contributions will yield preciser solutions, however, it is clear that the possible range of the fundamental mode will be confined to low percentages.\\
    
    \item \textbf{Light Sources:}\\
    The majority of applications require tightly focused light sources that have a high percentage in the fundamental mode. In addition to the FLs, diode lasers as well as lasers with gaseous gain media, such as Helium-Neon lasers and ion-lasers are important examples \cite{ST:2007}. In Table \ref{tab:my_label}, we allocate them to the respective $M^2$-regions. Typical $M^2$ values for light sources are given by the range $1.1 - 1.7$ \cite{ST:2007,Pecile:2024}, yielding $P_{00}$ approximations from $[0.9,0.95] - [0.3,0.65]$.\\
    
    \item \textbf{Optical resonators:}\\
    Coupling into optical resonators is governed by the mode-matching factor $\epsilon$, which describes the fraction of a beams' fundamental mode power \cite{Ye:2001}. As a result, there is a direct correspondence given by \begin{align}\epsilon = P_{00}\,.\end{align} This means that the results of the inequalities \eqref{awesome} and \eqref{awesome3} directly apply to the mode-matching factor. Hence, an $M^2$-measurement of the light source directly before the cavity can determine the range of coupling efficiency for the resonator.\\
    \end{small}
\end{itemize}
A clear takeaway from these examples is that for applications demanding high fundamental mode percentages above 90$\%$, $M_x^2+M_y^2< 2.2$ is a necessity, as can be seen in Fig. 2.

An alternative method to assess beam quality is through the use of optical ring cavities to do mode scans of the input beam and measure its HOM composition \cite{Kwee:2007}. We reference a publication from the Hannover Laser Zentrum e.V. \cite{Weßels:2003}, in which a pre-mode cleaner design, initially used for the LIGO project \cite{Willke:1998}, acts as a mode scanning device for a large-mode-area fiber amplifier. Their measurements for a continuous-wave source with $1064\, \si{\nano\meter}$ resulted in $P_{00}> 0.975$, as well as $M^2<1.05\pm 0.05$. Using our approximation in inequality \eqref{awesome3} on the $M^2$-value of 1.05, we find $P_{00}\in [0.95,0.975]$. This result is in good agreement with the mode scan measurement and covers the full possible range within the error margin. Compared to mode analysis with mode cleaners, an $M^2$-measurement yields a simpler way to assess the fundamental mode contribution.

\subsection{Conclusion}
In this work, we derived a generally applicable lower bound in inequality \eqref{awesome} as well as low-order approximated upper bound in inequality \eqref{awesome3} to the relative power of the fundamental mode in correlation with the simple means of an $M^2$-measurement. The inequality \begin{align}
   2-\frac{M_x^2+M_y^2}{2} \leq P_{00}\leq 1.5-\frac{M_x^2+M_y^2}{4}
\end{align} poses a novel and powerful tool that yields new applications for beam quality assessment via $M^2$.

The result of inequality \eqref{awesome3} will have high relevance for optical methods that benefit from fundamental mode operation, such as coupling into and out of optical resonators and fibers, long distance collimation and tight focusing. For applications with a strong HOM onset, e.g., high-power systems or multi-mode fibers, we present a mathematical tool through the equations \eqref{Param1}-\eqref{Param3} and \eqref{bound1}-\eqref{bound3} to derive the full range of $P_{00}$ up to any necessary higher order to achieve maximum precision fit to any experiments' specific needs.

We have demonstrated how the inequality framework \eqref{awesome} and \eqref{awesome3} enables direct mode purity estimation across a range of real-world systems -- including laser sources, fiber optics, and optical resonators. These examples represent just a subset of the broader applicability of $M^2$-based beam purity analysis, which stands to benefit diverse fields from atomic, molecular, optical and quantum physics to industrial photonics.

In conclusion, our work provides the first extensive method to extract mode purity from $M^2$-measurements. This result gives long-overdue clarity to a foundational concept in beam characterization -- placing the widely used statement “$M^2\approx 1$ implies high beam quality” on solid analytical ground.

\section{Methods}
In this section we provide a short overview of the mathematical methods used for the derivation of our main results given by equations \eqref{Param1}--\eqref{Param3}, inequality \eqref{awesome}, inequalities \eqref{spe1}--\eqref{spe3} and inequality \eqref{awesome3}. 

In principle, the most general parametrization in equations \eqref{Param1}--\eqref{Param3} is derived by solving the linear system of equations \eqref{sol1}--\eqref{sol3} with the assumption of at least 3 modes present in the beam. The detailed derivation, as well as the calculation of the inequalities \eqref{spe1}--\eqref{spe3} for the special cases, is provided in section 2 of the supplementary material. For a mode composition with less than 3 modes the approach is the same. However, additional conditions, further constraining the $M^2$-parameters, can be assumed. Inequality \eqref{awesome} directly follows from inequality \eqref{ineq1} through the reasoning given in section \ref{lowbou}.

The upper bound is derived by calculating the parameter space of $\lbrace P_{11},P_{20},P_{02}\rbrace$ by analyzing equations \eqref{bref1} and \eqref{bref2} for the boundary value problem given in equations \eqref{bound1}--\eqref{bound3}. The maximum of $P_{00}$ is then calculated on this parameter space by finding the maximum of $P_{11}+P_{20}+P_{02}$ on the boundary of the three-dimensional geometric object that corresponds to the parameter space, as discussed in section 3 of the supplementary material.

\section{Acknowledgements}
This research was funded in whole or in part by the Austrian Science Fund (FWF) [DOI: 10.55776/P36040
and 10.55776/F1004]. For open access purposes, the author has applied a CC by public copyright license to any author accepted manuscript version arising from this submission.

We also thank Lucile Rutkowski, Thomas Benoy and everyone from the FLAIR 2024 conference in Assisi/Italy for inspiring discussions during the poster session, as well as Markus Aspelmeyer and Garrett Cole for their valuable feedback on the manuscript.

\begin{landscape}
\begin{table}[]
    \centering
    \begin{tabular}{cccc}\hline
        $M^2$ & Lower bound & Upper bound & Examples\\\hline\\
        1.1 & 0.9 & 0.95 & fiber lasers (erbium, ytterbium, thulium), HeNe-lasers, single-mode fibers \cite{ST:2007}, \\
         & & & large-mode-area fiber amplifiers \cite{Weßels:2003} (all with $M^2<1.1$)\\
       \vdots & \vdots & \vdots &  \\\\
        1.3 & 0.7 & 0.85 & ion lasers ($M^2\in \left[1.1,1.3\right]$) \cite{ST:2007}, optical parametric oscillators in the mid-infrared (idler $M^2\in \left[1.2,1.3\right]$) \cite{Pecile:2024}\\\\
    \vdots & \vdots & \vdots &  \\\\
        1.7 & 0.3 & 0.65 & diode lasers ($M^2\in [1.1,1.7]$) \cite{ST:2007}\\\\
       \vdots & \vdots & \vdots &  \\\\
        2 & 0 & 0.5 & multi-mode fibers (arbitrary $M^2$-range possible, usually $M^2\geq 3$) \cite{ST:2007}\\\\\hline
    \end{tabular}
    \caption{This table is aimed at giving a good orientation of which $M^2$-values correspond to which values of fractional fundamental mode power with respective applications provided. We assume $M^2=M_x^2=M_y^2$ for representation purposes. To some of the values we assign examples whose $M^2$-values approximately fall within range of the corresponding scenario.}
    \label{tab:my_label}
\end{table}
\end{landscape}

\end{document}